\documentclass[epsf,prb,twocolumn,nofootinbib]{revtex4}

\usepackage[pdftex]{graphicx}
\usepackage{dcolumn}
\usepackage{bm}
\usepackage{epsfig}
\usepackage{latexsym}
\usepackage{amsmath}
\usepackage{amsfonts}
\usepackage{amssymb}
\usepackage{color}
\usepackage{array}
\usepackage{framed}

\newcommand{\oh}{{\frac{1}{2}}}

\newcommand{\xv}{{\bf x}}

\def\rf#1{(\ref{#1})}

\begin{document}

\title{Fracton-Elasticity Duality}
\author{Michael Pretko and Leo Radzihovsky}
\affiliation{
Department of Physics and Center for Theory of Quantum Matter\\
University of Colorado, Boulder, CO 80309}
\date{November 29, 2017}

\begin{abstract}
  Motivated by recent studies of fractons, we demonstrate that
  elasticity theory of a two-dimensional quantum crystal is dual to a
  fracton tensor gauge theory, providing a concrete manifestation of
  the fracton phenomenon in an ordinary solid.  The topological
  defects of elasticity theory map onto charges of the tensor gauge
  theory, with disclinations and dislocations corresponding to
  fractons and dipoles, respectively.  The transverse and longitudinal
  phonons of crystals map onto the two gapless gauge modes of the
  gauge theory.  The restricted dynamics of fractons matches with
  constraints on the mobility of lattice defects.  The duality leads
  to numerous predictions for phases and phase transitions of the
  fracton system, such as the existence of gauge theory counterparts
  to the (commensurate) crystal, supersolid, hexatic, and isotropic
  fluid phases of elasticity theory.  Extensions of this duality to
  generalized elasticity theories provide a route to the discovery of
  new fracton models.  As a further consequence, the duality implies
  that fracton phases are relevant to the study of interacting
  topological crystalline insulators.
\end{abstract}
\maketitle

\emph{Introduction}.  There has been a recent surge of theoretical
interest in a new class of quantum phases of matter featuring
excitations of restricted mobility.  The archetypal examples of this
phenomenon are models that exhibit ``fracton'' excitations, particles
which are strictly immobile in isolation, but which can move through
interaction with other particles.  More generally, there are particles
which move freely only along certain subspaces while being immobile in
the transverse directions, exhibiting subdimensional behavior.
Fractons and other subdimensional particles were first seen in the
context of certain exactly solvable lattice
models.\cite{chamon,bravyi,haah,cast,yoshida,haah2,fracton1,fracton2}
It was later realized that these exotic phases of matter have a
natural theoretical description in the language of tensor gauge
theories, which feature higher moment charge conservation laws
restricting the motion of particles.\cite{sub,genem,alex} There has
been rapid recent progress in the field, establishing connections with
quantum Hall systems,\cite{theta,matter} gravity,\cite{mach} and
glassy dynamics,\cite{abhinav,screening} among many other theoretical
developments.\cite{williamson,sagarlayer,hanlayer,parton,slagle,bowen,
  nonabel,balents,field,albert,correlation,simple,entanglement,bernevig}

\begin{figure}[t!]
 \centering
 \includegraphics[scale=0.44]{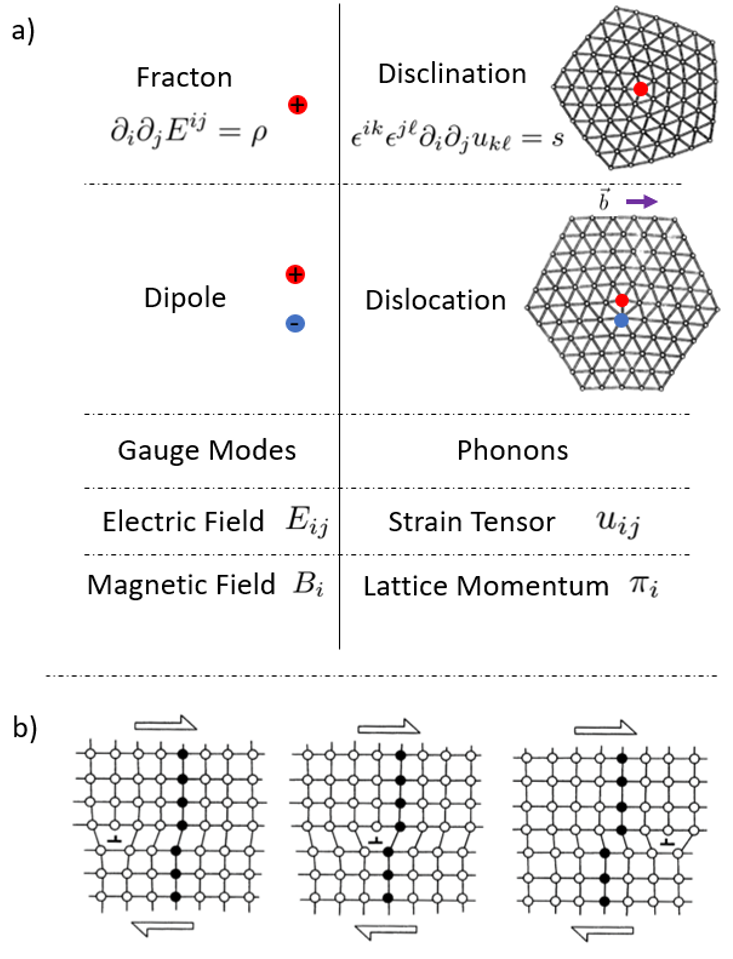}
 \caption{a) The Fracton-Elasticity Dictionary: Excitations and operators
   of the scalar-charge tensor-gauge theory are in one-to-one
   correspondence with those of a two-dimensional quantum crystal.
   (Pictures of lattice defects adapted from
   Ref. \onlinecite{seung}.)  b) A dislocation can only freely move by gliding along its Burgers vector $\vec{b}$, while dislocation climb (motion perpendicular to $\vec{b}$) requires the presence of vacancy/interstitial defects.}
 \label{fig:dictionary}
\end{figure}

Despite extensive studies of their exotic properties, fracton models
have so far been lacking concrete physical realizations.  To this end,
in this Letter we explicitly demonstrate that, intriguingly, a
two-dimensional quantum crystal realizes a fracton model described by
a noncompact rank-two tensor gauge theory.  This duality is a direct
tensor analogue of the familiar particle-vortex duality relating a
two-dimensional superfluid to a noncompact $U(1)$ gauge
theory.\cite{dasgupta,fisher} As summarized in
Fig.\ref{fig:dictionary}a, the longitudinal and transverse phonons of a
crystal map onto the two gapless gauge modes of the tensor gauge
theory, with the phonon momentum and strain tensor mapping onto the
magnetic and electric tensor fields.  Concomitantly, the topological
lattice defects map directly onto the gauge charges.  Specifically,
disclinations and dislocations correspond to fractons and dipoles,
respectively.  In this way, the constrained mobility of fracton models
is demystified in terms of the well-known restrictions on motion of a
crystal's topological defects.  Dislocations can only glide along their Burgers vector, as shown in Fig.\ref{fig:dictionary}b, while tranverse motion (dislocation
climb) requires the absorption or emission of vacancies and
interstitials.  Similarly, any motion of a disclination creates a
``scar'' of extra dislocations in the
crystal.\cite{emit1,emit2,emit3,emit4} There are no local processes
which move a single disclination, which is therefore immobile in
isolation, a manifestation of the fracton phenomenon.

Utilizing this duality, we make numerous predictions about the phases
and phase transitions of the fracton gauge theory by mapping onto
established results in elasticity theory.  For example, the fracton
system will exhibit natural gauge theory analogues of the commensurate
(vacancy/interstitial-free) crystal, supersolid, hexatic, and
isotropic fluid phases.  We can thereby also determine the critical
properties of transitions between these phases.  By generalizing the
duality to elasticity theories of other physical systems, such as
three-dimensional crystals, magnetic Wigner crystals and liquid
crystals, our arguments provide a route to the discovery of new
fracton phases.  In turn, the conservation laws of fracton gauge
theories provide a convenient and systematic tool for encoding and
analyzing the dynamics of crystal defects.  As a further application
of the duality, we discuss the relevance of fracton theories to the
study of interacting topological crystalline insulators (TCIs).

\emph{Duality}.  We begin by presenting a streamlined derivation
of fracton-elasticity duality, relegating a more detailed derivation
and discussion to a companion paper.\cite{forth} Dual gauge
formulations of elasticity theory have been investigated in the past
\cite{zaanen}, though with different focus and without making physical
connection with fracton theories, which is the aim of the present
work.  The theory of elasticity is conveniently formulated in terms of
a phonon vector field $u_i(\xv)$ representing the displacement of an
atom from its equilibrium position.  The low-energy action for the
displacement is given by\cite{chaikin,landau,kardar},
\begin{equation}
S = \int d^2xdt\,\frac{1}{2}\left[(\partial_t u^i)^2 -
C^{ijk\ell}u_{ij}u_{k\ell}\right],
\label{orig}
\end{equation}
where $u_{ij} = \frac{1}{2}(\partial_i u_j + \partial_j u_i)$ is the
linear part of the symmetric strain tensor and $C^{ijk\ell}$ is a
matrix of elastic constants, with its components determined by the
underlying lattice.  It is useful to separate the displacement field
into its singular and smooth phonon pieces, in terms of which we write
$u_{ij} = u_{ij}^{(s)} + \frac{1}{2}(\partial_i\tilde{u}_j
+ \partial_j\tilde{u}_i)$, where $\tilde{u}_i$ is a smooth
single-valued function, and the singular strain component
$u_{ij}^{(s)}$ is sourced by topological defects via
\begin{equation}
  \epsilon^{ik}\epsilon^{j\ell}\partial_i\partial_ju_{k\ell} =
 \epsilon^{ik}\epsilon^{j\ell}\partial_i\partial_ju^{(s)}_{k\ell} = s.
\label{disc}
\end{equation} 
The disclination density $s =
\epsilon^{ij}\partial_i\partial_j\theta_b$ is defined as a singularity
of the bond angle, $\theta_b=\epsilon^{k\ell}\partial_ku_\ell$, giving
$s =
\epsilon^{ij}\partial_i\partial_j(\epsilon^{k\ell}\partial_ku_\ell)$.
Dislocations are also implicitly accounted for in this treatment,
since a dislocation can be regarded as a bound state of two
disclinations,\cite{chaikin,landau,seung,sarang} as we will see explicitly.

We now introduce two Hubbard-Stratonovich fields, a momentum vector
$\pi_i$ and a symmetric stress tensor $\sigma_{ij}$.  In terms of
these variables, we rewrite the action as,
\begin{align}
\begin{split}
S = \int d^2x dt\,&\bigg[\frac{1}{2}C^{-1}_{ijk\ell}\sigma^{ij}\sigma^{k\ell} - \frac{1}{2}\pi^i\pi_i \\
& - \sigma^{ij}(\partial_i \tilde{u}_j + u_{ij}^{(s)}) + \pi^i\partial_t (\tilde{u}_i + u_i^{(s)}) \bigg],\label{action2} 
\end{split}
\end{align}
with the original action recovered by integrating out the fields
$\pi_i$ and $\sigma_{ij}$.  The smooth displacement field
$\tilde{u}_i$ can now be integrated out, thereby enforcing the
constraint,
\begin{equation}
\partial_t \pi^i - \partial_j\sigma^{ij} = 0,\label{Newton}
\end{equation}
which is simply the continuum form of the Newton's equation of motion,
relating the stress imbalance to the rate of change of lattice
momentum.  To solve this constraint it is convenient to first
introduce rotated field redefinitions, $B^i = \epsilon^{ij}\pi_j$ and
$E_\sigma^{ij} = \epsilon^{ik}\epsilon^{j\ell}\sigma_{k\ell}$, which
transforms the Newton equation constraint (\ref{Newton}) into the
generalized Faraday equation, appearing in fracton tensor gauge
theories,\cite{genem}
\begin{equation}
\partial_t B^i + \epsilon_{jk}\partial^j E_\sigma^{ki} = 0.
\end{equation}
The label $\sigma$ on the field $E_\sigma^{ij}$ indicates its relation
to the rotated stress tensor. 

The general solution to this equation is conveniently expressed in
terms of a symmetric rank-2 tensor gauge field, $A^{ij}$, and a scalar
potential, $\phi$, (in analogy with the potential formulation of
Maxwell's vector electrodynamics)
\begin{equation}
  B^i = \epsilon_{jk}\partial^j A^{ki}\,,\,\,\,\,\,\,\,\,\,
E_\sigma^{ij} = -\partial_t A^{ij} - \partial_i\partial_j\phi\;,
\label{potentials}
\end{equation}
with $\phi$ playing the role of the Airy stress function of static
elasticity theory.  Note that the fields $E_\sigma^{ij}$ and $B^i$
are invariant under the generalized gauge transformation on the
potentials,
\begin{equation}
A_{ij}\rightarrow A_{ij}+\partial_i\partial_j\alpha\,,
\,\,\,\,\,\,\,\,\,\,\,\,\,\,\,\,\,\phi\rightarrow\phi+\partial_t\alpha
\end{equation}
for arbitrary function $\alpha(\xv,t)$.  The potential formulation has
therefore introduced a gauge redundancy into the problem. Expressing
the action (\ref{action2}) in terms of electric and magnetic fields,
using the potentials in (\ref{potentials}) inside the last two terms,
integrating by parts and utilizing the definition of the disclination
density (\ref{disc}), we obtain,
\begin{align}
\begin{split}
  S = \int
  d^2xdt\left[\frac{1}{2}\tilde{C}^{-1}_{ijk\ell}E_\sigma^{ij}E_\sigma^{k\ell}
    - \frac{1}{2}B^iB_i -\rho\phi - J^{ij}A_{ij}\right],
\end{split}
\label{finalact}
\end{align}
where $\tilde{C}^{ijk\ell} =
\epsilon^{ia}\epsilon^{jb}\epsilon^{kc}\epsilon^{\ell d}C_{abcd}$ is a
function of the elastic coefficients, $\rho = s$ is the disclination
density, and $J^{ij} =
\epsilon^{ik}\epsilon^{j\ell}(\partial_t\partial_k
- \partial_k\partial_t)u_\ell$ is the
current tensor capturing the motion of disclocations and disclinations, as introduced in Ref.\;\onlinecite{MRvortices,genem}.  For a dislocation with Burgers vector $b^\ell$ moving at velocity $v^j$, this tensor takes the form $J^{ij} = \epsilon^{(i\ell}v^{j)}b_\ell$,\cite{foot5} with the trace $J^i_{\,\,i}$ describing dislocation climb.\cite{MRvortices,zaanen2}  The action of Eq.\ref{finalact} is in precisely the form of the
scalar-charge tensor gauge theory, allowing for anisotropy in the
electric field term, with disclinations playing the role of fracton
charges.\cite{genem,theta} This action leads to two gapless gauge
modes, corresponding to the longitudinal and transverse phonon modes
of elasticity theory.

We note in passing that this gauge theory does not support instanton events in the path integral, which would correspond to terms in the elasticity
Hamiltonian which explicitly break translational symmetry and gap out
the phonon modes, as could arise via coupling to a substrate.  For a conventional crystal, which breaks an
underlying translational symmetry {\em spontaneously}, instantons are
forbidden and the gauge field is noncompact, as discussed further in
the companion paper.\cite{forth}

It will also be useful to introduce a canonical conjugate electric
tensor field, $E_{ij} = -\partial\mathcal{L}/\partial \dot{A}^{ij} =
\tilde{C}^{-1}_{ijk\ell}E_\sigma^{k\ell}$, in terms of which the
tensor gauge theory Hamiltonian is given by
\begin{equation}
H = \int d^2x\,\bigg(\frac{1}{2}\tilde{C}^{ijk\ell}E_{ij}E_{k\ell} + \frac{1}{2}B^iB_i + \rho\phi + J^{ij}A_{ij}\bigg).
\end{equation}
Note that the scalar potential $\phi$ does not have a conjugate field,
but rather acts as a Lagrange multiplier enforcing the scalar Gauss's
law constraint,
\begin{equation}
\partial_i\partial_j E^{ij} = \rho.
\label{gauss}
\end{equation}
This constraint is the dual formulation of Eq.\rf{disc}, defining the
disclination density.  We see that the duality maps $E^{ij}$ to a
rotated strain tensor via $E^{ij} =
\epsilon^{ik}\epsilon^{j\ell}u_{k\ell}$, while the closely-related
``velocity''-like field, $E_\sigma^{ij}$ is mapped to a rotated stress
tensor via $E_\sigma^{ij} =
\epsilon^{ik}\epsilon^{j\ell}\sigma_{k\ell}$.  The relation
$E_\sigma^{ij} = \tilde{C}^{ijk\ell}E_{k\ell}$ between the two
electric field tensors exactly mirrors the relation $\sigma^{ij} =
C^{ijk\ell}u_{k\ell}$ between the stress and strain tensors.  The
Gauss's law (\ref{gauss}) is notable for leading to conservation of
both charge and dipole moment\cite{sub}
\begin{equation}
Q = \int d^2x\,\rho=\textrm{const.}\,,\,\,\,\,\,\,\,\,{\bf P} = \int
d^2x\,(\rho\xv) = \textrm{const.}
\end{equation}
The conservation of dipole moment has the dramatic consequence that an
isolated charge is strictly locked in place, since a motion of a
fracton charge proceeds by a creation of a dipole moment, and thus
would violate dipole charge conservation.  The presence of this extra
conservation law therefore directly encodes the fractonic behavior of
disclinations.

The dipole moment conservation law also implies that a dipole is a
topologically stable excitation, since it cannot decay directly into
the vacuum.  In elasticity language, this corresponds to a bound state
of two equal and opposite disclinations, known as the dislocation
defect.\cite{chaikin,landau} We can check this correspondence
explicitly by studying the total dipole moment contained in a region
$V$.  Assuming the region has zero net charge (so that dipole moment
is independent of origin), we can write the dipole moment in the form,
\begin{align}
\begin{split}
P^i &= \int_V d^2x\,(\rho x^i) =  \oint_{\partial V} ds^j\partial_j(\epsilon^{ik}u_k),\\
&= \epsilon^{ik}\Delta u_k = \epsilon^{ik}b_k\ ,
\end{split}
\end{align}
where $\Delta u_k$ is the net change in the displacement $u_k$ going
around the boundary of a region $V$, which is precisely the definition
of a Burgers vector ${\bf b}$.  From this, we see that the dipole
matches explicitly with a dislocation defect, ${\bf P}=\hat{\bf
  z}\times{\bf b}$, with the dipole vector perpendicular to the
Burgers vector.  With this correspondence in place, the
fracton-elasticity dictionary is now complete, as summarized in
Fig.\ref{fig:dictionary}.

One important additional property of a crystal that dual gauge theory
must capture is that in the absence of vacancies and interstitials a
dislocation can only move along its Burgers vector, $i.e.$, can glide
but is unable to climb.  On the other hand, by itself a conservation
of a dipole moment does not place any fundamental restriction on the
motion of a dipole. To see how the one-dimensional constrained dipole
dynamics arises in the tensor gauge theory, we consider a particular
component of the quadrupole moment.  Following the standard analysis
of fracton theories,\cite{sub} it is straightforward to derive the
following conservation law,
\begin{equation}
\int d^2x\,(\rho x^2 - 2E^i_{\,\,i}) = \textrm{const.}
\end{equation}
Any longitudinal motion of a dipole requires a change of this
quadrupole moment, which, as we see from the above constraint is
necessarily accompanied by a change in $E^i_{\,\,i}$.  To understand
the physical meaning of this conservation law, we rewrite the trace in
elasticity language,
\begin{equation}
E^i_{\,\,i} = \partial_i u^i = n_d + \partial_i\tilde{u}^i,
\end{equation}
where we have broken up the divergence into $n_d$, the number of
vacancies minus the number of interstitial defects, and a smooth
elastic piece $\tilde{u}^i$.  We can then write our conservation law
as,
\begin{equation}
\int d^2x\,(\rho x^2 - 2n_d) = \textrm{const.}
\end{equation}
In other words, the longitudinal motion of a dipole (corresponding to
a dislocation climb) requires the absorption or creation of vacancies
or interstitial defects.  This provides a kinetic and energetic
barrier, which, in the absence of vacancies and interstitials
constrain dislocations and the corresponding fracton dipoles into
quasi-one-dimensional particles, as expected.

To see more formally and explicitly that longitudinal dipole motion
(equivalently, dislocation motion transverse to its Burgers
vector, $i.e.$, a climb) creates vacancy/interstitial defects, we
examine the Ampere equation of motion, $\delta H/\delta A_{ij} = 0$,
which takes the form,
\begin{equation}
\partial_t E^{ij} + \frac{1}{2}(\epsilon^{ik}\partial_kB^j 
+ \epsilon^{jk}\partial_kB^i) = -J^{ij}.
\end{equation}
The piece of this equation that is relevant for our purposes is
the trace, which can be written as\cite{MRvortices}
\begin{equation}
\partial_t n_d + \partial_iJ^i_d = -J^i_{\,\,i}\ ,
\label{continuity}
\end{equation}
where $J^i_d = \pi^i = \epsilon^{ji}B_j$ is the current density of
vacancies and interstitials, and we have used the fact that
$E^i_{\,\,i}\approx n_d$, since $\partial_i \tilde{u}^i\ll 1$.  The
above equation represents a continuity equation for the
vacancy/interstitial number, sourced by a dislocation current
transverse to the Burgers vector, $J^i_{\,\,i}=\hat{\bf z}\cdot({\bf v}\times{\bf b})$,
indicating that dislocation climb creates vacancy/interstitial
defects.\cite{MRvortices}

\begin{figure}[t!]
 \centering
 \includegraphics[scale=0.17]{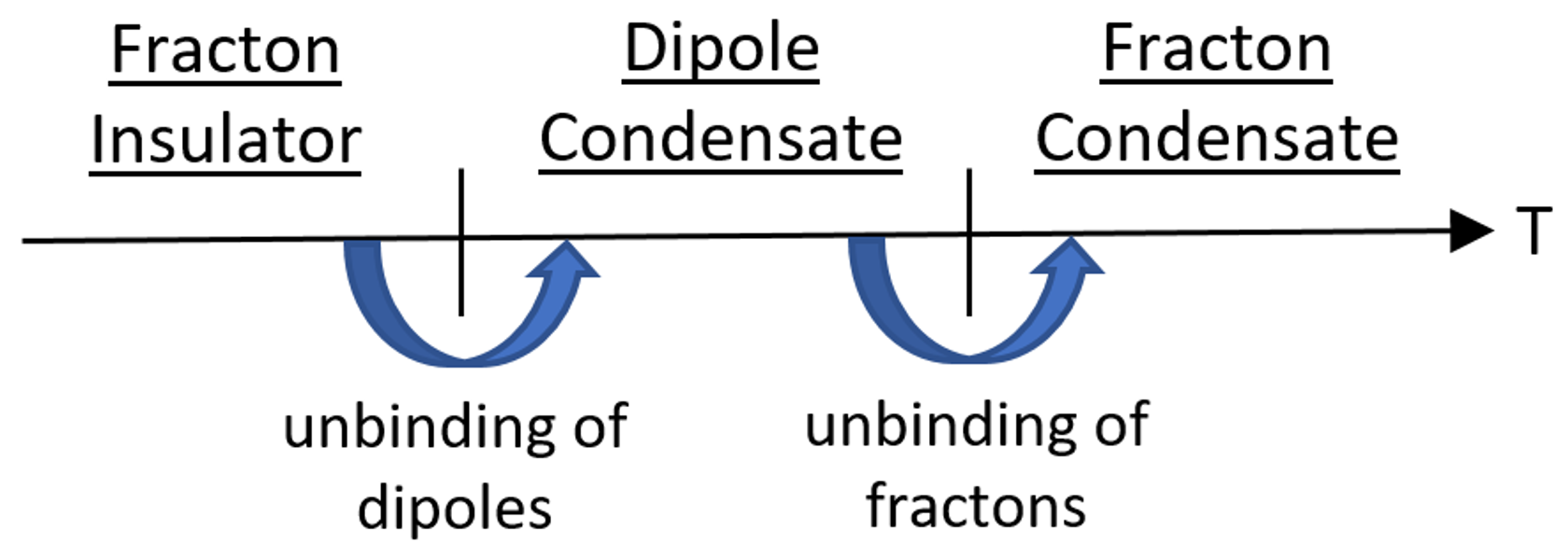}
 \caption{The duality with elasticity theory predicts that the fracton gauge theory exhibits two finite-temperature phase transitions, corresponding to unbinding of dipoles and fractons, respectively.}
\label{fig:transitions}
\end{figure}

\emph{Dual fracton superconductor}. The duality has mapped a 2d
crystal onto a rank-two gauge theory coupled to fracton matter, with the
dual gauge theory action \rf{finalact} naturally describing a fracton insulator. However, to access finite density fracton phases,
it is convenient to introduce more explicit coupling to matter fields\cite{forth}
\begin{eqnarray}
  S &=& \int d^2xdt\bigg[\frac{1}{2}\tilde{C}^{-1}_{ijk\ell}
E_\sigma^{ij}E_\sigma^{k\ell}
- \frac{1}{2}B^iB_i\nonumber\\ 
&& - \oh g_1(\partial_i\partial_j\theta - A_{ij})^2 
+ \oh g_0(\partial_t\theta - \phi)^2\bigg],
\label{SCaction}
\end{eqnarray}
where $\theta$ is the phase of the fracton field and the $g_i$ are
determined by core energies of the defects.  This action is capable of
describing a dual fracton ``superconductor" ($i.e.$, a condensate of
fractons), with the normal phase ($i.e.$, the fracton insulator)
corresponding to the crystal.  This action also supports a third
phase between the fracton superconductor and insulator, as we discuss
next.

\emph{Applications}.  The field of fractons is still in the early
stages of development, and thus lacks much of the basic machinery used
in the study of symmetry breaking systems and conventional topological
phases.  As such, much less is known about the various phases and
phase transitions of fracton models (though recent progress has been
made on this subject).\cite{matter} For the specific fracton model
discussed here, however, we can obtain the entire phase diagram and
characterize the nature of phase transitions by the above mapping onto
a two-dimensional crystal, which has been studied in great
detail.\cite{halperin,nelson,young} The duality thereby gives key
features of phases and phase transitions of the above scalar charge fracton
model, which we expect to also provide insight into more general
fracton systems.

More specifically, in addition to the above established correspondence
between a crystal and gauged fracton insulator, two fracton-proliferated phases emerge as
duals of the orientationally ordered ($e.g.$, hexatic) and isotropic
fluids.  On the elasticity side, these appear at finite temperature as a result of two-stage BKT-like melting transitions: (i) a
crystal-to-hexatic fluid transition, at which dislocations (that are
logarithmically bound in a crystal)
proliferate,\cite{halperin,nelson,young}, followed by (ii) a
hexatic-to-isotropic fluid BKT transition\cite{ber1,ber2,kt}, at which
disclinations (bound quadratically in the crystal phase, but screened
down to logarithmic binding in the hexatic) entropically proliferate.
We thus predict a finite temperature fracton phase diagram with three distinct phases, distinguished by the proliferation of dipoles and fractons, as summarized in Fig.\ref{fig:transitions}.  The proliferated phases can be regarded as a dipole condensate and a fracton condensate, respectively, with implications for the quantum theory of melting, discussed elsewhere.\cite{forth}  These transitions are all
captured by the tensor dual ``superconductor'' model, \rf{SCaction},
that at finite temperature reduces to a classical 2d tensor
sine-Gordon model. We leave the more detailed analysis of these
fracton phases and transitions on the gauge theory side to future
research.\cite{forth}

We also note that at zero temperature two qualitatively distinct
quantum crystal phases are allowed. A ``commensurate crystal'' (with
the weight of Bragg peaks commensurate with the number of particles)
is characterized by long-range positional and orientational orders and
a vacuum of gapped vacancies and interstitials, $i.e.$, a Mott
insulator. With increased quantum fluctuations ($e.g.$, reduced mass),
vacancies and interstitials condense at finite density into an
``incommensurate crystal'', that is a
supersolid\cite{Andreev,MosesChen,ketterle} in the case of bosonic
atoms. The fracton-elasticity duality thus predicts two distinct
zero-temperature fracton insulating phases on the tensor gauge theory
side, distinguished by gapped and condensed quadrupole excitations. 

We conclude by noting that fracton-elasticity duality draws an intriguing
connection to a seemingly unrelated subject of classification of
interacting crystal symmetry protected topological insulators
(TCIs).\cite{kane,bernevig2,konig,fu,tireview,
  tci,tcireview,senthil,xie,chong,chong2} In classifying interacting
symmetry-protected topological (SPTs) phases, one particularly
powerful tool is gauging the symmetry protecting the SPT
phase.\cite{gauging} The result is a topologically ordered
state\cite{foot4}, described by a gauge theory with a gauge group
equivalent to the symmetry group of the original SPT phase, with
different interacting SPT phases corresponding to distinct
topological phases.

For internal symmetry groups, this gauging procedure is fairly
straightforward, done by coupling to a dynamical flux of the symmetry
group.  However, for the case of spatial symmetries, the notion of
flux insertion is less clear.  As recently demonstrated\cite{else},
flux of a crystal symmetry is equivalent to a lattice defect, with
a dislocation and a disclination respectively corresponding to a flux
of translational and rotational symmetries.  A resulting model with a
fully gauged crystalline symmetry exhibits dynamical lattice defects,
i.e., it is a quantum elasticity theory.  Fracton-elasticity duality
then allows us to map the gauged system onto a fracton theory.  Hence,
the result of gauging a two-dimensional crystalline symmetry is a
fracton phase, as opposed to the more conventional topological
phases obtained by gauging an internal symmetry.  We expect that a
more detailed understanding of fracton phases thus obtained by gauging
crystal symmetries may prove useful for classifying interacting TCIs,
a quest that is still being actively
pursued.\cite{hiroki,hao,sj1,meng,sungjoon,zou,sj2} We leave the
details of implementing this program as a task for the future.

\emph{Conclusions}.  In this manuscript, we have explicitly
demonstrated a duality between two-dimensional quantum elasticity and
a fracton tensor gauge theory, in a natural tensor generalization of
conventional particle-vortex duality.  The topological defects of a 2d
crystal map directly onto the charges and dipoles of the gauge theory,
while phonons and elastic strain tensor respectively correspond to the
gapless gauge modes and the tensor electric field.  This duality
demystifies the constrained mobility of fractons and dipoles by
mapping them onto known properties of disclinations and dislocations,
respectively.  As a result, we made predictions about fracton phases
and phase transitions by mapping onto the phase diagram of quantum
crystals.  Our work opens the door for the future exchange of ideas
between the emerging field of fractons and the well-established study
of elasticity theory.

\emph{Acknowledgments}.  The authors acknowledge useful conversations
with Yang-Zhi Chou, Han Ma, Matthew Fisher, Shriya Pai, Abhinav Prem,
Albert Schmitz, Rahul Nandkishore, and Mike Hermele. This work was
supported by the Simons Investigator Award to LR from the Simons
Foundation and partially by the NSF Grant 1734006.  LR thanks the KITP for its hospitality as part of the Fall 2016 Synthetic Matter workshop and sabbatical program, during which this work was initiated and supported by the NSF grant PHY-1125915.  LR was also supported by the Soft Materials Research Center 
under NSF MRSEC Grants DMR-1420736.


\begin{thebibliography}{9}
\bibitem{foot5}
This tensor will also have a contribution from processes in which a disclination (fracton) moves while emitting or absorbing a dislocation (dipole).
\bibitem{foot4} 
More precisely, the system will have long-range
  entanglement.  For a discrete symmetry group, the resulting gauge
  theory will be gapped, and we can rigorously use the term
  topological order.  For a continuous symmetry group, however, the
  resulting gauge theory will be gapless.
\end{thebibliography}
\end{document}